\documentclass{PoS}

\usepackage{graphicx}
\usepackage{amsmath,amssymb}
\usepackage{psfrag}
\usepackage{capt-of}
\usepackage{bm}
\newcommand{\ba}{\begin{eqnarray}}
\newcommand{\ea}{\end{eqnarray}}
\newcommand{\be} {\begin{equation}}
\newcommand{\ee} {\end{equation}}

\newcommand{\bra}[1]{\ensuremath{\langle{#1}|}}
\newcommand{\ket}[1]{\ensuremath{|{#1}\rangle}}
\newcommand{\mr}[1]{\ensuremath{\mathrm{#1}}}

\newcommand{\mthb}[1]{\ensuremath{\mathbf{#1}}}
\title{Semileptonic decays of $K$ and $D$ mesons in $2+1$ flavor QCD}

\ShortTitle{Semileptonic decays of $K$ and $D$ mesons in $2+1$ flavor QCD}

\author{%
Jon~A.~Bailey$^a$
A.~Bazavov$^b$,
C.~Bernard$^c$,
C.~Bouchard$^{a,d}$,
C.~DeTar$^e$,
A.X.~El-Khadra$^d$,
E.D.~Freeland$^c$,
\speaker{E.~G\'amiz}$^{a}$,
Steven~Gottlieb$^{f,g}$,
U.M.~Heller$^h$,
J.E.~Hetrick$^i$,
A.S.~Kronfeld$^a$,
J.~Laiho$^j$,
L.~Levkova$^d$,
P.B.~Mackenzie$^a$,
M.B.~Oktay$^d$,
J.N.~Simone$^a$,
R.~Sugar$^k$,
D.~Toussaint$^b$,
and
R.S.~Van~de~Water$^l$ \\ \\
\llap{$^a$}Fermi National Accelerator Laboratory,\hspace*{-0.4em}
    \thanks{Operated by Fermi Research Alliance, LLC, under Contract
    No.~DE-AC02-07CH11359 with the United States Department of Energy.}~
 Batavia, IL, USA \\
\llap{$^b$}Department of Physics, University of Arizona, Tucson, AZ  
85721, USA \\
\llap{$^c$}Department of Physics, Washington University, St.~Louis, MO  
63130, USA \\
\llap{$^d$}Physics Department, University of Illinois, Urbana, IL  61801, 
USA \\
\llap{$^e$}Physics Department, University of Utah, Salt Lake City, UT  
84112, USA \\
\llap{$^f$}Department of Physics, Indiana University, Bloomington, IN  47405, 
USA \\
\llap{$^g$}National Center for Supercomputing Applications, University of 
Illinois, Urbana, IL  61801, USA \\
\llap{$^h$}American Physical Society, One Research Road, Ridge, NY  11961, 
USA \\
\llap{$^i$}Physics Department, University of the Pacific, Stockton, CA  95211, 
USA \\
\llap{$^j$}SUPA, School of Physics \& Astronomy, University of Glasgow, 
Glasgow, G12 8QQ, UK\\
\llap{$^k$}Department of Physics, University of California, Santa Barbara, 
CA  93106, USA \\
\llap{$^l$}Department of Physics, Brookhaven National 
Laboratory,\hspace*{-0.4em}
    \thanks{Operated by Brookhaven Science Associates, LLC, under 
    Contract No.\ DE-AC02-98CH10886 with the United States Department 
    of Energy.}~
  Upton, NY, USA \\

Email: \email{egamiz@fnal.gov}}
\author{Fermilab Lattice and MILC Collaborations\\
}

\abstract{The experimentally measured rates of the semileptonic decays 
$K \to \pi l \nu$ and $D \to K(\pi) l \nu$ can be combined with lattice 
calculations of the associated form factors to precisely extract the CKM 
matrix elements $|V_{us}|$ and $|V_{cs(d)}|$. We report on the status of 
form factor calculations with Fermilab charm quarks and staggered light 
quarks on the 2+1 flavor asqtad staggered MILC ensembles. Analysis of
data for the $D\to\pi l\nu$ form factor provides a nontrivial test of our methods
via comparison with 
CLEO data.  We discuss the use of HISQ valence quarks to 
calculate the $K \to \pi l \nu$ form factor $f_+^{K\pi}(0)$ and describe 
tests of our method.
}

\FullConference{The XXVIII International Symposium on Lattice Field Theory, 
Lattice2010\\
		June 14-19, 2010\\
		Villasimius, Italy}

\begin{document}

\section{Introduction}

\label{introduccion}

Studies of exclusive semileptonic decays of $B$, $D$, and $K$ mesons are 
used to extract the Cabibbo-Kobayashi-Maskawa (CKM) matrix elements $|V_{ub}|$, $|V_{cb}|$, $|V_{cs}|$, $|V_{cd}|$, and $|V_{us}|$ with errors 
competitive with those obtained using inclusive 
semileptonic decays, leptonic decays, neutrino-antineutrino interactions, and 
$\tau$ decays~\cite{ReviewRuthLat09}.  The theory inputs needed to 
fix the CKM matrix elements from exclusive semileptonic widths are 
form factors parameterizing corresponding hadronic matrix elements:
\ba \label{formfac}
\langle P_2\vert V^\mu\vert P_1\rangle  = f_+^{P_1P_2}(q^2)(p_{P_1}+
p_{P_2}-\Delta)^\mu +f_0^{P_1P_2}(q^2)\Delta^\mu\,,
\ea
where  $\Delta^\mu=(m_{P_1}^2-m_{P_2}^2)q^\mu/q^2$, 
$q=p_{P_1}-p_{P_2}$, and $V$ is the appropriate flavor-changing vector current.
Alternatively we may write \cite{ElKhadra:2001rv}
\begin{equation}
\bra{P_2}V_\mu\ket{P_1}=\sqrt{2m_{P_1}}[v_\mu f_\Vert^{P_1P_2}(q^2)
+p_{\bot\mu}f_\bot^{P_1P_2}(q^2)],
\end{equation}
where $v=p_{P_1}/m_{P_1}$ and $p_\bot=p_{P_2}-(v\cdot p_{P_2})v$, so that 
in the rest frame of a heavy meson $P_1$,
\begin{equation}
f_\Vert^{P_1P_2}(q^2)=\frac{\bra{P_2}V^0\ket{P_1}}{\sqrt{2m_{P_1}}}\quad\mathrm{and}
\quad f_\bot^{P_1P_2}(q^2)=\frac{\bra{P_2}V^i\ket{P_1}}{\sqrt{2m_{P_1}}}\frac{1}
{p_{P_2}^i}.
\end{equation}

Typically, theoretical errors in the form factors
limit the accuracy of such extractions of the CKM matrix elements. The situation
has been acute in the case of $D$ semileptonic decays~\cite{Besson:2009uv}.
Here we describe calculations of $D$ and $K$ semileptonic form factors, which provide access to $|V_{cs(d)}|$ and $|V_{us}|$, respectively.

For $D$ decays we seek not only the CKM matrix elements, but also to validate 
applying our methods to the $B$ decays $B\to\pi l \nu$ and 
$B\to K l \bar l$. Below we use a subset
of the available lattice data to check our methods;
we compare the shape of a preliminary result for the $D\to\pi l\nu$ form factor
with the shape as measured by CLEO~\cite{Besson:2009uv}.
For $D$ decays, unlike $B$ decays,
the lattice and experimental data overlap throughout most of the $q^2$ domain, affording a more powerful check.

Precise determinations of $\vert V_{us}\vert$ 
provide stringent tests of first-row unitarity and may furnish
additional information about the scale of new physics~\cite{exptf+}.  
Here we describe the main ingredients of our strategy to use
staggered quarks to obtain $f_+^{K\pi}(0)$
and the tests we have performed to verify that our approach will 
yield errors competitive with existing calculations of the form factor.

\section{$D\to\pi l \nu$: Extraction of $\vert V_{cd}\vert$}

\label{sec:Dtopi}

\subsection{\label{subsec:ens}Ensembles and valence masses}
We have completed generating correlators with Fermilab heavy quarks and asqtad staggered light quarks on the 2+1 flavor asqtad staggered MILC ensembles shown in Table~\ref{table:ens}.  The heavy quark is tuned to the charm mass on each ensemble, and the light valence masses include partially quenched and full QCD points.
In addition to the ensembles shown in Table~\ref{table:ens}, we are generating correlators on a fine ensemble with $m_l=0.15m_s$, superfine ensembles with $m_l\approx0.14m_s,\ 0.1m_s$, and an ultrafine ($a\approx0.045\ \mr{fm}$) ensemble with $m_l=0.2m_s$.
However, the analysis presented below is restricted to full QCD data from the coarse $0.4m_s$ and $0.2m_s$ ensembles and the fine ensembles shown in Table~\ref{table:ens}.

\begin{table}[tbp]
\hspace*{-0.5cm}
\vspace*{-0.5cm}
\caption{\label{table:ens}MILC ensembles~\cite{Bazavov:2009bb,Bernard:2001av,Aubin:2004wf} for the current round of $D\to\pi(K)l\nu$ analyses, together with the valence masses used for all ensembles at each lattice spacing.  Valence masses after the semicolons are the tuned strange mass.  Data generation is complete for all ensembles and quark masses shown.}
\vspace*{0.2cm}
\centering
\begin{tabular}{lcccclc}
\hline\hline
       & $\approx a$ (fm) & $am_l/am_s$ & $N_s^3\times N_t$ & $N_{conf}$ & $am_\mr{valence}$ \\
\hline
coarse & $0.12$ & $0.02/0.05$ & $20^3\times64$ & $2052$ & $0.005,\ 0.007,\ 0.01,$ \\ 
        &       & $0.01/0.05$ & $20^3\times64$ & $2259$ & $0.02,\ 0.03,\ 0.0415,$ \\ 
        &       & $0.007/0.05$ & $20^3\times64$ & $2110$ & $0.05;\ 0.0349$        \\
        &       & $0.005/0.05$ & $24^3\times64$ & $2099$ &                        \\
\hline
fine   & $0.09$ & $0.0124/0.031$ & $28^3\times96$ & $1996$ & $0.0031,\ 0.0047,\ 0.0062,$\\
        &       & $0.0062/0.031$ & $28^3\times96$ & $1946$ & $0.0093,\ 0.0124,\ 0.031;$  \\
        &       & $0.0031/0.031$ & $40^3\times96$ & $1015$ & $0.0261$                     \\
\hline
superfine & $0.06$ & $0.0072/0.018$ & $48^3\times144$ & $593$ & $0.0036,\ 0.0072,\ 0.0018,$ \\
          &        & $0.0036/0.018$ & $48^3\times144$ & $668$ & $0.0025,\ 0.0054,\ 0.0160;$ \\
          &        &                &                 &       & $0.0188$                    \\
\hline\hline
\end{tabular}
\hspace*{-0.5cm}
\end{table}

\subsection{\label{corrs}Correlators, correlator ratios, and ratio fits}

\begin{center}
\begin{figure}[b]
\begin{minipage}[c]{0.48\textwidth}
\begin{center}
\vspace*{0.5cm}
\hspace*{-0.2cm}\includegraphics[width=1.\textwidth]{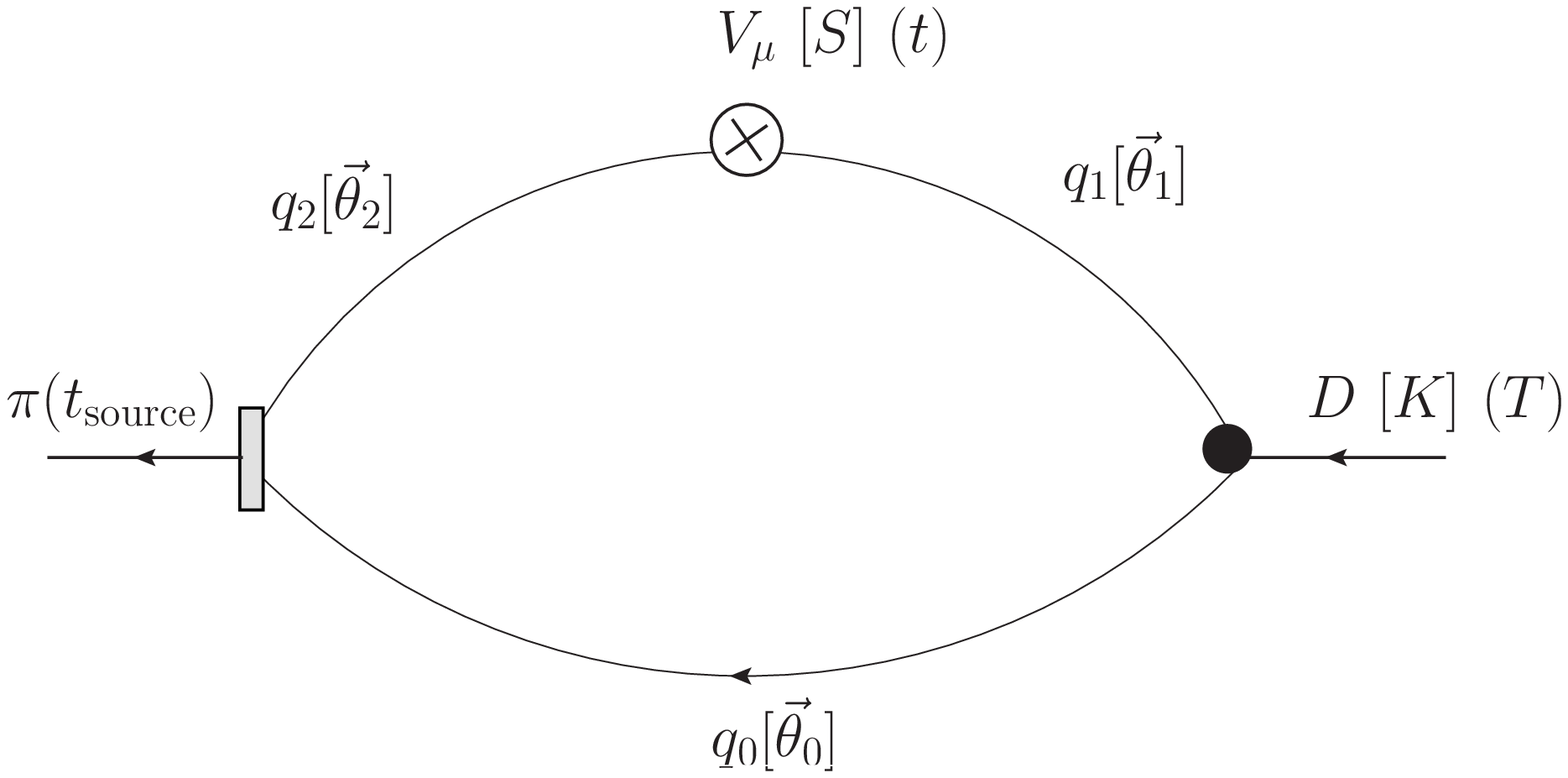}
\end{center}
\captionof{figure}{Structure of the 3-point functions needed to 
calculate $f_{\Vert,\bot}^{D\pi}[f_0^{K\pi}]$. Light quark propagators are 
generated at $t_{source}$ with local sources [random wall sources]. An 
extended charm [strange] propagator is generated at $T$. 
\label{fig:diagram}}
\vspace*{-0.5cm}
\end{minipage}
\hspace*{0.3cm}
\begin{minipage}[c]{0.48\textwidth}
\begin{center}
\psfrag{R-par-plat}[b][b][0.7]{$R_\Vert\sim f_\Vert^{D\pi}$}
\psfrag{current operator time slice}[b][b][0.7]{$V_4$ insertion time $t$}
{
\vspace*{-0.15cm}\includegraphics[width=0.8\textwidth]{./Dibujos/plateau_v2.eps}
}
\captionof{figure}{Fits of preliminary results for $f_\Vert^{D\pi}$ 
to plateaus and excited-state exponentials. Errors are statistical only and were obtained from 500 bootstrap ensembles. 
\label{fig:ex-interp_fvert}}
\end{center}
\vspace*{-0.5cm}
\end{minipage}
\end{figure}
\end{center}

To extract the matrix elements $\bra{\pi}V_\mu\ket{D}$ corresponding to 
$f_{\Vert,\bot}^{D\pi}(q^2)$, we use ratios of 3-point to 2-point correlators designed to cancel oscillations of opposite-parity states in staggered correlators~\cite{Btopi08}.
To minimize statistical errors and avoid excited-state contamination, we generate the 3-point correlators at two source-sink separations~\cite{Bailey:2009pz}.  The structure of the 3-point correlators is shown in Fig.~\ref{fig:diagram}.
The 3-point correlators are computed with current insertions at all times between the source and sink.  For insertion times far from the source and sink, plateaus appear in the ratios.  These plateaus are proportional to the desired form factors $f_\Vert^{D\pi}$ and $f_\bot^{D\pi}$.

\vspace*{-0.9cm}
We are studying combinations of fit ranges,
fit functions, source-sink--separations, and momenta to minimize
errors and control excited-state contamination.  
Fits to ratios yielding $f_\Vert^{D\pi}$ on the coarse $0.1m_s$ ensemble are 
shown in Fig.~\ref{fig:ex-interp_fvert};
we fit to a constant with an exponential on 
the $D$-side of the 3-point correlator to account for leading excited-state contributions; the resulting curves are consistent with the data.  In Fig.~\ref{fig:ex-interp_fvert} we also plot the resulting plateau terms and bootstrap errors over the entire fit ranges.
Consistent with
expectations~\cite{Bailey:2009pz}, we find the larger source-sink separation 
is optimal for $\mthb{p}=\mthb{0}$, while the smaller source-sink separation is 
optimal for $|\mthb{p}|>0$.

\hspace*{-0.8cm}
\begin{figure}
\vspace*{-0.8cm}
\begin{minipage}[c]{0.48\textwidth}
\begin{center}
\hspace{1.5mm}
\psfrag{r1-1/2fperp}[b][b][0.7]{$r_1^{-1/2}f_\bot^{D\pi}$}
\psfrag{r1E^2}[b][b][0.6]{$(r_1E_\pi)^2$}
{
\vspace*{-0.6cm}
\includegraphics[width=0.8\textwidth]{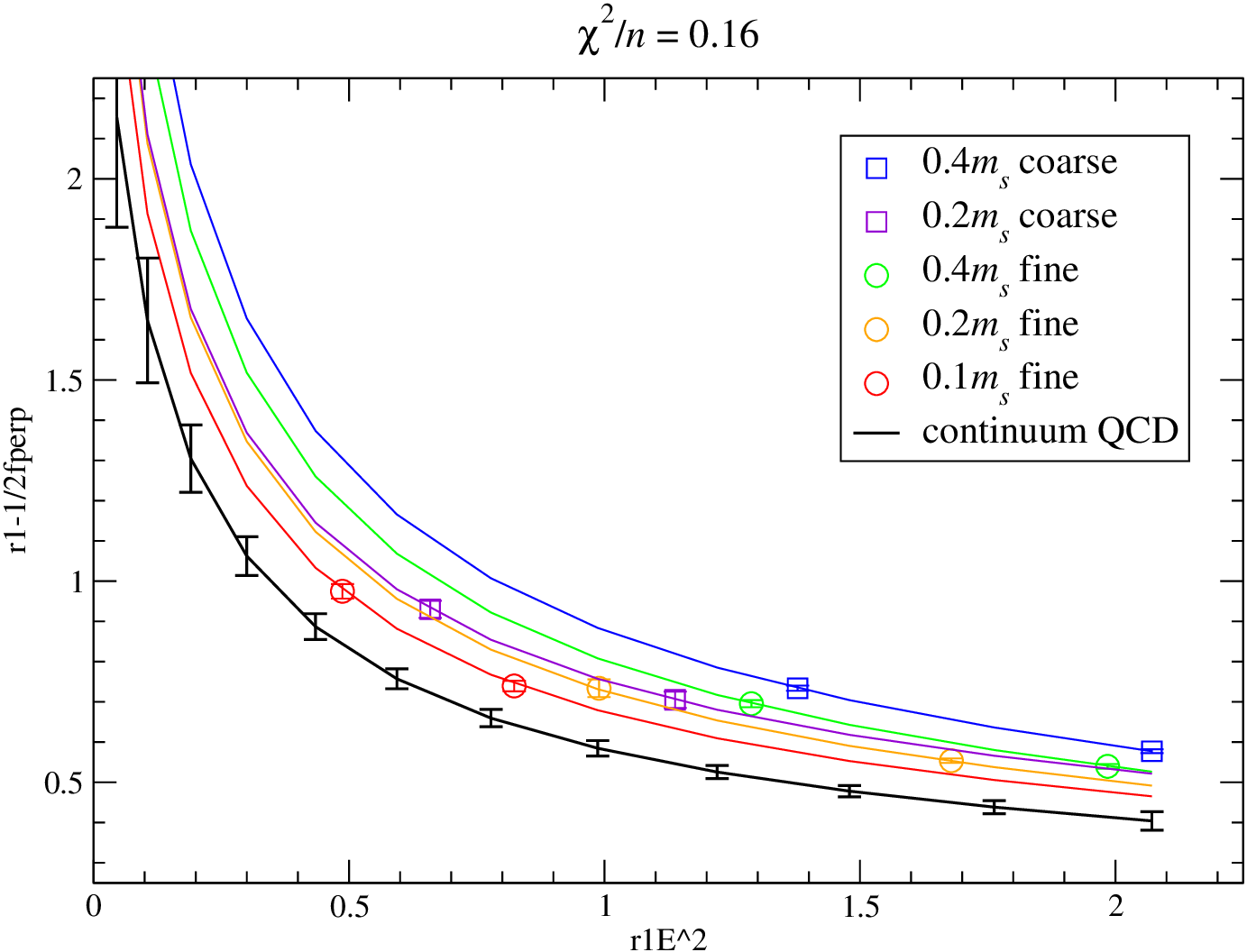}
}
\vspace*{0.7cm}
\captionof{figure}{A simultaneous fit to S$\chi$PT of all $f_\bot^{D\pi}$ data from the indicated ensembles.  
Errors are statistical and were obtained with 500 bootstrap ensembles. 
The black curve is the continuum result at physical quark masses and 
fiducial energies.
\label{fig:ex-interp_bot}}
\vspace*{0.4cm}
\end{center}
\end{minipage}
\hspace*{0.3cm}
\begin{minipage}[c]{0.48\textwidth}
\begin{center}
\vspace*{0.5cm}
\includegraphics[width=0.8\textwidth]{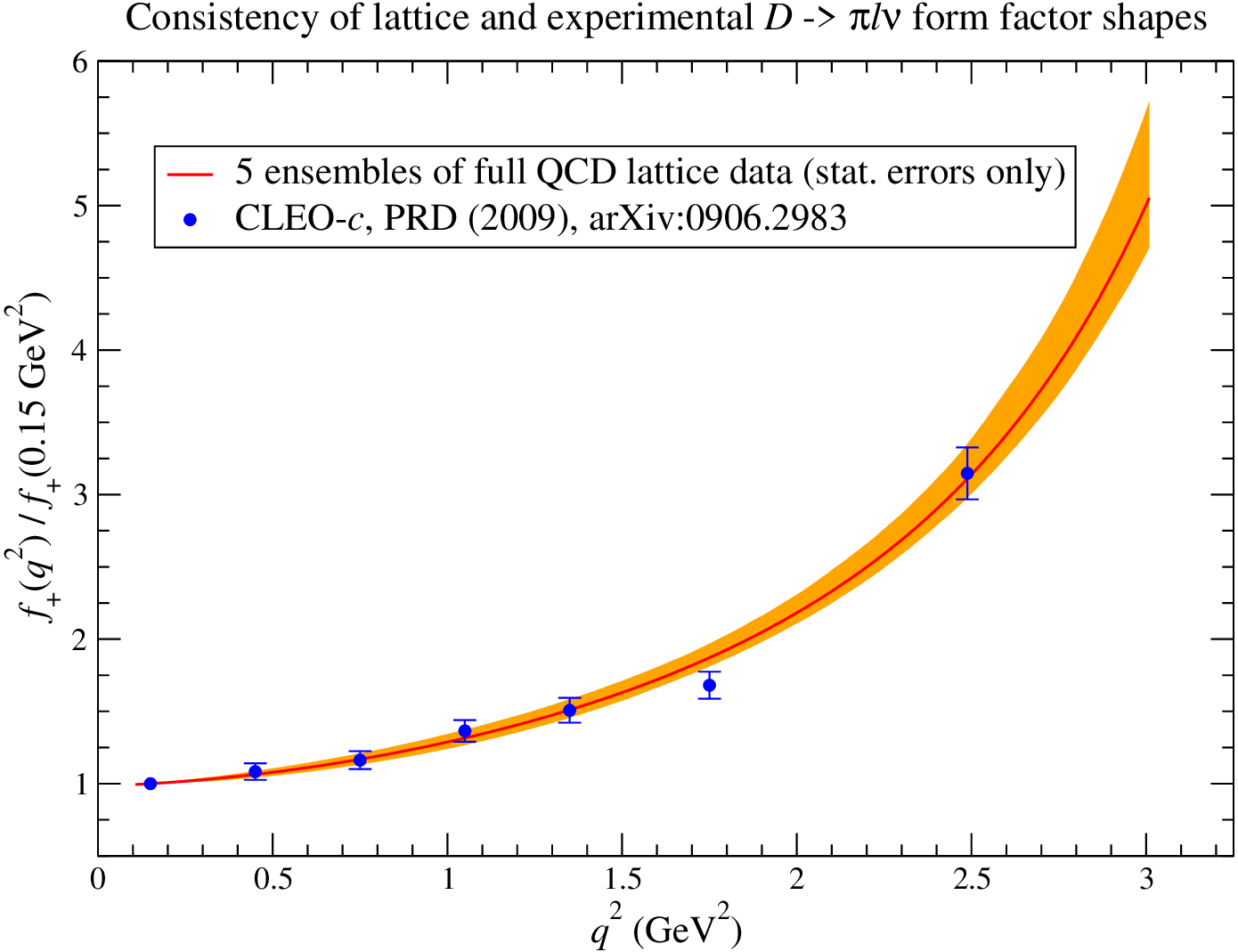}
\captionof{figure}{Overlay of the ratio $f_+^{D\pi}(q^2)/f_+^{D\pi}
({\tilde q}^2)$ from the lattice (red curve and orange error band) 
and CLEO (blue points).  The orange error band shows statistical lattice 
errors, and the blue error bars, the full experimental errors.
At ${\tilde q}^2=0.15\ \mr{GeV^2}$, the 
results agree and the errors vanish by definition.
\label{fig:check}}
\end{center}
\end{minipage}
\vspace*{-0.5cm}
\end{figure}

\subsection{Renormalization and blinding}

We need to renormalize the current. This provides an easy way to do a 
blind analysis. After nonperturbatively renormalizing the quark fields in the current, the remaining lattice artifacts at leading order in HQET are perturbatively calculable.  A subset of our collaboration is calculating this correction, which enters the result as an overall multiplicative factor depending on the ensemble and valence masses.  By including an offset in this factor, the normalization of the form factors and the implied value of $|V_{cd}|$ is masked from analysts performing fits, thereby eliminating a potential source of bias.

\subsection{Chiral-continuum-energy extra-interpolation}

We fit the form factors $f_{\Vert,\bot}^{D\pi}$ obtained from the correlator ratios to NLO heavy-meson S$\chi$PT~\cite{Aubin:2007mc} and extrapolate the results to the physical quark masses and continuum limit.  For the comparison below of lattice and experimental results, we also used S$\chi$PT to describe the energy dependence of the form factor, and we supplement the NLO expressions with NNLO terms analytic in the quark masses and lattice spacing.
Although we include data from only a subset of the ensembles (cf.~Sec.~\ref{subsec:ens}), we perform a simultaneous fit to all data from all ensembles included.
We include momenta through $\frac{2\pi}{aN_s}(1,1,0)$ and obtain statistical errors by propagating the bootstrap errors from the ratio fits.
A fit to the data for $f_\bot^{D\pi}$ is shown in Fig.~\ref{fig:ex-interp_bot}.  The results are stable under variations of the prior central values and addition of NNNLO analytic terms.

\subsection{Comparison of lattice and experimental form factor shapes}
To compare our result with experiment, we consider the ratio $f_+^{D\pi}(q^2)/f_+^{D\pi}({\tilde q}^2)$, where ${\tilde q}^2\equiv0.15\ \mr{GeV^2}$ is a convenient but otherwise arbitrary reference point.  The ratio $f_+^{D\pi}(q^2)/f_+^{D\pi}({\tilde q}^2)$ can be fixed from experiment without the CKM matrix element $|V_{cd}|$.  Using this ratio to compare the shapes of the lattice and experimental results also cancels the blinding factor. The use of this type of ratio to compare lattice and experimental results was advocated in \cite{visualization}.

In Fig.~\ref{fig:check}, we overlay the (preliminary) lattice and (currently final) experimental results for the ratio $f_+^{D\pi}(q^2)/f_+^{D\pi}({\tilde q}^2)$.
The red curve shows the lattice central value, and the orange error band shows the bootstrap errors.  The blue data points show the experimental central values, and the blue error bars, the statistical and systematic errors from the full covariance matrix~\cite{Besson:2009uv}.

\subsection{$D\to\pi l\nu$: Summary and next steps}

The shape of our preliminary result for the $D\to\pi l\nu$ form factor, obtained from a subset of our data, closely matches the shape seen in the CLEO data.  
This agreement encourages us to apply our methods to the calculations of the 
$B$ semileptonic form factors and related searches for new physics.
The statistical errors in the lattice form factor at the fiducial value ${\tilde q}^2$ are about $5\%$, in accord with expectations~\cite{Bailey:2009pz}.
We are adding to the analysis partially quenched and full QCD data from the remaining two coarse ensembles in Table~\ref{table:ens} and the superfine $0.4m_s$ and $0.2m_s$ ensembles.  Estimates of heavy quark errors, the uncertainty propagated from the $D^*D\pi$ coupling, and other systematics are in progress.

Finally, we are exploring combining information about the energy-dependence of the form factors from the $z$-expansion with information about the quark mass and lattice spacing dependence from S$\chi$PT by using S$\chi$PT to compute the mass and lattice spacing dependence of the Taylor coefficients in the $z$-expansion.  This approach furnishes an alternative to S$\chi$PT for model-independent, simultaneous fits of data at all energies on all ensembles, and is similar to, but distinct from, that detailed in~\cite{Na:2010uf}.

\section{$K\to\pi l\nu$: Exploring methodology to simulate at $q^2=0$}

One of the most significant systematic errors in traditional lattice 
analyses of $K\to\pi l \nu$ arises because
correlation functions with periodic boundary conditions do not cover 
the physical region of $q^2$, so obtaining
$f_+^{K\pi}(q^2=0)$ and extracting $\vert V_{us}\vert$ from 
experimental data requires interpolating between $q^2_{max}$ and 
unphysical values of $q^2$. 
Model dependence is introduced by the choice of interpolating function.
We want to eliminate this systematic error by using twisted boundary 
conditions to simulate at $q^2\simeq 0$. This approach was first suggested 
in \cite{FJS05} and later exploited with 2+1 flavors of domain wall fermions 
in \cite{RBCKtopi09} and  2 flavors of twisted mass fermions 
in \cite{ETMCKtopi09}.

The other main component of our analysis is the method 
developed by the HPQCD Collaboration to study $D$ semileptonic decays 
\cite{HPQCD_Dtopi}.  This method is based on the Ward identity
relating the matrix element of a vector current to that of the 
corresponding scalar current 
$q^\mu\langle \pi\vert V_\mu^{lat.} \vert K\rangle Z=(m_s-m_q) 
\langle \pi\vert S^{lat.} \vert K\rangle\,$, 
with $S=\bar s l$, and $Z$, a lattice renormalization factor for the vector 
current.  Using the definition of the form factors in Eq.~(\ref{formfac}) 
and this identity, one can extract
$f_0^{K\pi}(q^2)$ at any $q^2$
by using
\ba
f_0^{K\pi}(q^2) = \frac{m_s-m_l}{m_K^2-m_\pi^2}\langle \pi\vert S 
\vert K\rangle (q^2).
\ea
The kinematic constraint requires $f_+(0)=f_0(0)$, so this relation can be 
used to calculate $f_+^{K\pi}(0)$. The downside of the method is that it 
gives no access to the shape of $f_+^{K\pi}$, but that is
very well known from experiment \cite{exptf+}. For more details of this 
method, see \cite{HPQCD_Dtopi}.

\subsection{Test run: simulation and fitting details}

The main goal of this test run is a realistic estimate 
of the statistical errors we could achieve 
and an assessment of how easily we can tune the twisting angles to get 
values of $q^2$ close to zero. 
For these tests we used about 500--600 configurations from each of the coarse 
$0.2m_s$ and $0.4m_s$ ensembles, and $\sim550$ configurations for the fine 
$0.2m_s$ ensemble.  Instead of using the asqtad action for 
the light and strange valence quarks, as in our calculation 
of the $D\to\pi l\nu$ form factor, we use the HISQ formulation~\cite{Hisq}, 
which has better control of discretization effects.

\begin{figure}[b]
\begin{minipage}[c]{0.47\textwidth}
\begin{center}
\begin{tabular}{cccc}
\hline\hline
$\vert\bm{\theta}_1\vert$ & $\vert\bm{\theta}_2\vert$ & 
$(r_1q)^2$ \\
\hline
0 & 0 & 0.0227(3)\\
{\bf 0} & {\bf 0.7295} & {\bf 0.0011(4)}\\
0.7295 & 0 & 0.0153(3)\\
0 & 0.9105 & -0.0109(5)\\
0.9105 & 0 & 0.0114(5)\\
{\bf 1.2876} & {\bf 0} & {\bf 0.0003(3)}\\ 
\hline\hline
\end{tabular}
\end{center}

{\small 
\captionof{table}{\label{table:mom}Simulation values of the twisting angles 
$\bm{\theta}_1$ and 
$\bm{\theta}_2$, and the corresponding $q^2$. Errors are 
statistical. The smallest $q^2$ available with 
periodic boundary conditions is $\simeq -0.104$, 
already outside the physical region. Lines in bold correspond to 
$q^2\simeq0$.}
}
\end{minipage}
\hspace*{0.3cm}
\begin{minipage}[c]{0.48\textwidth}
\begin{center}
\vspace*{-0.9cm}\includegraphics[width=0.75\textwidth,angle=-90]{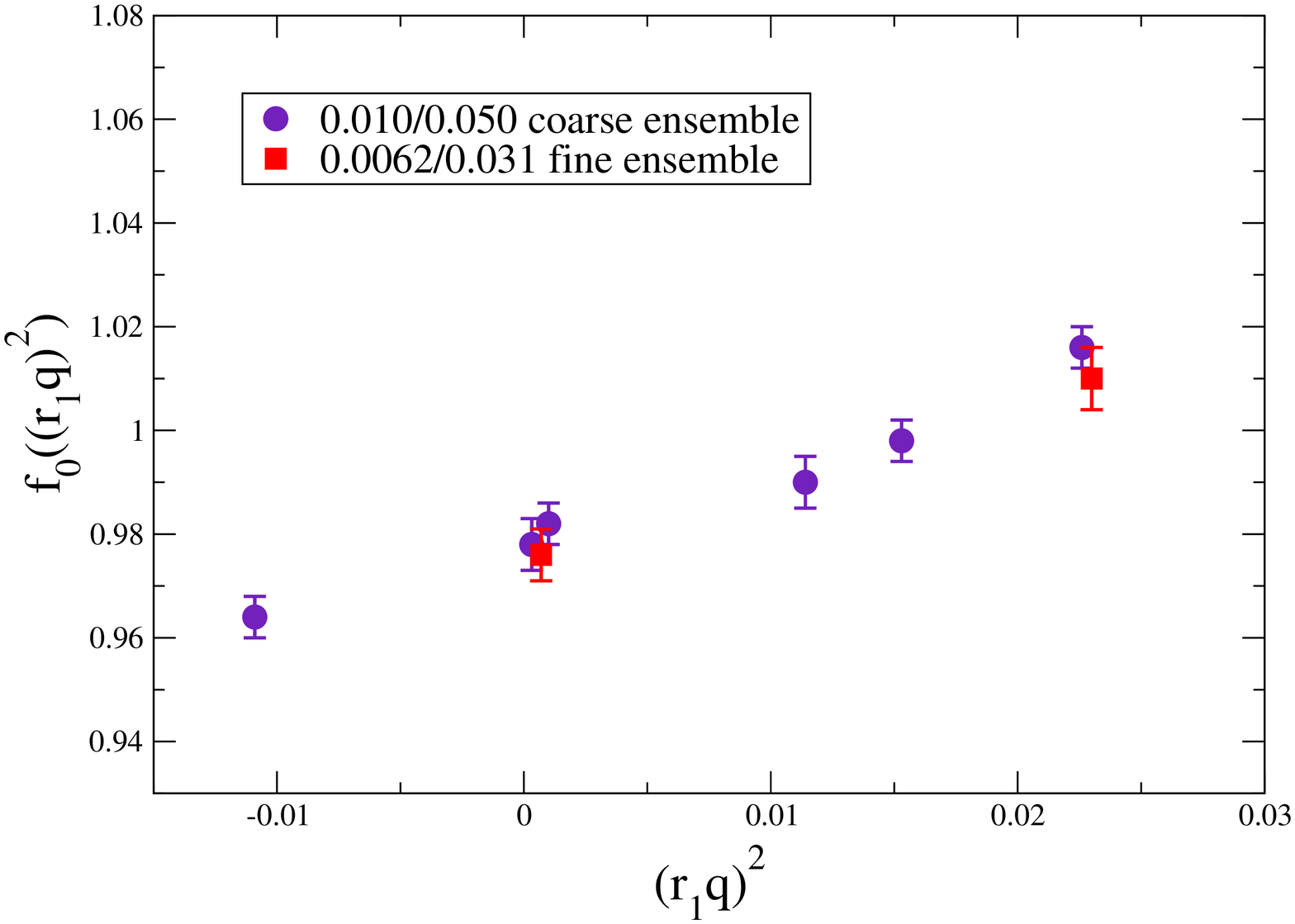}
\vspace*{-0.2cm}
\captionof{figure}{Form factor $f_0^{K\pi}$ as a function of momentum transfer
$q^2$ for the coarse and fine lattice points. \label{fig:resultscoarse}}
\end{center}
\vspace*{0.6cm}
\end{minipage}
\vspace*{-0.6cm}
\end{figure}

We generate 3-point correlators as shown in Fig.~\ref{fig:diagram}
with a scalar insertion at time $t$ and 2-point functions for 
kaons and pions using both local and random wall sources. The latter sources 
produce results with statistical errors 
2--3 times smaller than the former, so in the following discussion
we consider only the results obtained with random wall 
sources. We inject momentum in the 2-point functions by using twisted 
boundary conditions to generate (one of) the light propagators. For the 
3-point functions, we inject the momentum ${\bf p}=\bm{\theta}\pi/L$ in 
either the kaon or the pion 
by choosing either $\bm{\theta}_0=\bm{\theta}_2=\bm{0}$, 
$\bm{\theta}_1\ne\bm{0}$ or $\bm{\theta}_0=
\bm{\theta}_1=\bm{0}$, $\bm{\theta}_2\ne\bm{0}$, respectively 
(see Fig.~\ref{fig:diagram}). The different 
external momenta and resulting $q^2$ are shown in 
Table~\ref{table:mom}. We have obtained two values of $q^2$ very close to zero by 
tuning the twisting angle from 2-point correlator fits only. 
To extract the form factor, we fit the 3-point 
and 2-point correlators together, which gives us slightly different values 
for $q^2$, but still close enough to zero to avoid any significant interpolation in 
$q^2$. In fact, the values of $f_+^{K\pi}(q^2\simeq 0)$ 
that we obtain from the correlators with external momentum 
injected in the kaon and the pion agree within one
sigma.

We repeat the combined fits using iterative averages of the correlators at
different values of $t$ and $T$ to suppress oscillations due to opposite-parity states~\cite{Btopi08},
including 
the ground state and first oscillating contributions. We use fits to 
the ground state alone and fits including four
exponentials to crosscheck the central values and errors.

\subsection{Test results and future plans}

The results for the $f_0^{K\pi}(q^2)$ form factor for the different values 
of $q^2$ simulated on the $0.2m_s$ coarse and fine lattices are shown in 
Fig.~\ref{fig:resultscoarse}. The statistical errors for the two coarse points 
and the fine point with $q^2\simeq 0$ are about $0.4\%$. 
In Fig.~\ref{fig:resultscoarse} one can see that the form factors for 
$a\simeq 0.12,0.09\,{\rm fm}$ agree with each other within statistics, suggesting 
very small discretization effects. Similar behavior is observed when 
comparing results with $m_l=0.2m_s$ and $m_l=0.4m_s$ on the 
coarse lattices. Such behavior suggests that, after 
extrapolation to the continuum and the physical sea light quark masses, 
residual effects for those error sources will be negligible.

Using the full statistics available in these ensembles, 
around 4 times the number of configurations used here, 
we expect statistical errors around $0.2-0.3\%$.
Since we will eliminate the uncertainty due to the $q^2$ interpolation by 
simulating at $q^2\simeq 0$, the only 
significant remaining error besides statistics will be the one associated 
with the chiral-continuum extrapolation. We plan to do this
extrapolation using continuum $\chi$PT at NNLO and incorporate taste-breaking 
effects at NLO, including data from at least three different lattice spacings 
and light quark masses down to $m_s/8$.  
Based on the tests described here, we expect 
our calculation to be competitive with the existing state-of-the-art calculations 
of $f_+^{K\pi}(0)$ \cite{RBCKtopi09,ETMCKtopi09}.

Computations for this work were carried out in part on facilities of
the USQCD Collaboration, which are funded by the Office of Science of
the U.S. Department of Energy.
This work was supported in part by the U.S. Department of Energy under 
Grants No.~DE-FC02-06ER41446 (C.D., L.L., M.B.O),
No.~DE-FG02-91ER40661 (S.G.), No.~DE-FG02-91ER40677 (C.M.B., A.X.K., E.D.F.),
No.~DE-FG02-91ER40628 (C.B, E.D.F.), No.~DE-FG02-04ER-41298 (D.T.); 
the National Science Foundation under Grants No.~PHY-0555243, No.~PHY-0757333,
No.~PHY-0703296 (C.D., L.L., M.B.O), No.~PHY-0757035 (R.S.), 
No.~PHY-0704171 (J.E.H.) and No.~PHY-0555235 (E.D.F.).
C.M.B. was supported in part by a Fermilab Fellowship in Theoretical 
Physics
and by the Visiting Scholars Program of Universities Research Association, Inc.
R.S.V. acknowledges support from BNL via the Goldhaber Distinguished 
Fellowship.

\end{document}